\documentclass[twocolumn,showpacs,preprintnumbers,amsmath,amssymb,prb,aps]{revtex4}

\usepackage{epsfig}
\usepackage{graphicx}
\usepackage{dcolumn}
\usepackage{bm}

\begin{document}

\title{Electronic and magnetic properties of a quasi-one-dimensional spin
chain system, Sr$_{3}$NiRhO$_{6}$}

\author{S. K. Pandey and Kalobaran Maiti}

\altaffiliation{Electronic mail: kbmaiti@tifr.res.in}

\affiliation{Department of Condensed Matter Physics and Materials
Science, Tata Institute of Fundamental Research, Homi Bhabha Road,
Colaba, Mumbai - 400 005, INDIA}

\date{\today}

\begin{abstract}
We investigate the electronic structure of Sr$_3$NiRhO$_6$, a
quasi-one-dimensional spin chain system using \emph{ab initio} band
structure calculations. Spin polarized calculations within GGA
reveal that Ni and Rh have finite moments and they are
antiferromagnetically coupled along the chain axis in the ground
state. While these results obtained within the local spin density
approximations provide remarkable representation of the magnetic
phase, the experimentally observed insulating behavior could not be
captured within this method. GGA+$U$ calculations show that opening
up of an insulating gap requires on-site Coulomb interaction among
the Rh 4$d$ electrons, $U_{dd}^{Rh}$ $\approx$ 2.5 eV and the
correlation among Ni 3$d$ electrons, $U_{dd}^{Ni}$ $\approx$ 4.5 eV
suggesting this system to be a Mott insulator. Electron correlation
among $d$ electrons leads to significant enhancement of the O 2$p$
character in the energy bands in the vicinity of the Fermi level and
the $d$ bands appear at lower energies. Energy gap in the up spin
density of states appears to be significantly small ($\sim$ 0.12 eV)
while it is $>$ 2 eV in the down spin density of states suggesting
possibility of spin polarized conduction in the semiconducting
phase.
\end{abstract}


\maketitle

\section{Introduction}

Recently, quasi-one-dimensional compounds, $AE_3MM^\prime$O$_6$
($AE$ = alkaline earth metals such as Ca, Sr {\it etc.}; $M$ and
$M^\prime$ are transition metals) possessing rhombohedral
K$_4$CdCl$_6$ structure (space group \emph{R$\overline{3}$C}) have
attracted a great deal of attention due to their fascinating
magnetic
properties.\cite{niitakajssc,niitakaprl,sampath,stitzer,sengupta,flahaut,whangbo,vidya,sampath1,fresard,wuprl,
takubo,niharika,wuprb,sugiyama,agrestini} In this structure (see
Fig. 1), $M$O$_6$ forms in trigonal prismatic geometry and
$M^\prime$O$_6$ forms in octahedral geometry. These two units appear
alternatively along the chain direction, $c$-axis and are connected
via face sharing as shown in the figure. It is evident that the
interchain interaction is significantly weak making these compounds
a quasi-one-dimensional system. It is observed that
antiferromagnetic, ferromagnetic or ferrimagnetic long-range ordered
phases can be achieved by tuning the intrachain magnetic coupling in
these one dimension chains.\cite{niitakaprl,niharika,wuprb} In
addition to such interesting properties of the quasi-one-dimensional
chains, the whole system can be viewed as made off antiferromagnetic
or ferromagnetic chains arranged in a triangular lattice. Numerous
studies based on tailoring the composition of these geometrically
frustrated systems reveal plethora of novel phases such as spin
liquid phase, partially disordered antiferromagnetic phase {\it
etc.}\cite{sampath,flahaut,agrestini,niitakaprl,niharika}

In this paper, we report our results on the electronic structure of
Sr$_3$NiRhO$_6$ using full potential linearized augmented plane wave
method. This compound exhibits ferrimagnetic intrachain ordering
below 45 K. Further lowering in temperature leads to partially
disordered antiferromagnetic phase below 10 K.\cite{niharika}
Interestingly, analogous compound Sr$_3$NiPtO$_6$ exhibits
spin-liquid behavior. While it is difficult to capture disorder
effect and/or spin liquid phase using such {\it ab initio} band
structure calculations, Sr$_3$NiRhO$_6$ is a good starting point to
understand the intrachain coupling and associated ground state
properties.

Our results clearly establish that the ground state of this compound
consists of magnetic Ni and Rh ions which are antiferromagnetically
coupled along the $c$-axis. To capture the insulating transport
consistent with experimental observations, one needs to consider the
on-site Coulomb interaction among Ni 3$d$ and Rh 4$d$ electrons in
addition to the intrachain antiferromagnetic (IAFM) coupling, which
suggests that this compound is a Mott insulator. On-site Coulomb
interaction among Rh 4$d$ electrons is found to be lower (about
60\%) than that found for Ni 3$d$ electrons as expected due to
larger radial extension of the Rh 4$d$ orbitals.

\section{Computational details}

The spin polarized GGA (generalized gradient approximation) and
GGA+$U$ ($U$ = on-site Coulomb repulsion strength) electronic
structure calculations were carried out using {\it state-of-the art}
full potential linearized augmented plane wave (FPLAPW)
method.\cite{blaha} The lattice parameters and atomic positions used
in the calculations are taken from literature.\cite{stitzer} The
Muffin-Tin sphere radii were chosen to be 2.33, 2.19, 2.01, and 1.78
a.u. for Sr, Ni, Rh, and O atoms, respectively. For the exchange
correlation functional, we have adopted recently developed GGA form
of Wu {\em et al.}\cite{wu} The on-site Coulomb interactions were
considered within LSDA+$U$ (LSDA = local spin density approximation)
formulation of Anisimov {\em et al}.\cite{anisimov93} The
calculations were performed by varying on-site Coulomb interaction
among Ni 3$d$ electrons, $U_{dd}^{Ni}$ and Rh 4$d$ electrons,
$U_{dd}^{Rh}$. The convergence was achieved by considering 512 $k$
points within the first Brillouin zone and the error bar for the
energy convergence was set to be smaller than 10$^{-4}$ Ryd/cell.

\section{Results and discussions}

The crystal structure of Sr$_{3}$NiRhO$_{6}$ is hexagonal as shown
in Fig. 1, where the Sr, Ni, Rh, and O ions are denoted by spheres
of decreasing size. It is evident from the figure that the Ni and Rh
ions are surrounded by six oxygen ions forming trigonal prism and
octahedron, respectively. The trigonal prisms and octahedra are
connected by face sharing and form the chains along the $c$-axis. If
the axis system is defined such that $z$-axis is along $c$-axis and
$x$- and $y$-axis are in the $ab$-plane, {\it both} Ni 3$d$ and Rh
4$d$ orbitals defined in this axis system will not be degenerate due
to the corresponding crystal field effect. There will be three bands
consisting of $d_0$, $d_{\pm1}$ and $d_{\pm2}$ orbitals. The energy
separation and the location of the bands in the energy axis will
depend on the type of crystal field is applicable as discussed later
in the text.

\begin{figure}
  \includegraphics[width=10cm]{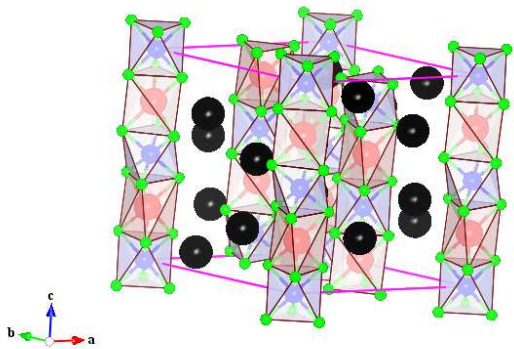}\\
  \vspace*{-3.5cm}
  \caption{(Color online) The unit cell of Sr$_3$NiRhO$_6$. Sr is
represented by largest symbols (black). Gradually decreasing sized
symbols represent Ni (pink), Rh (blue) and O (green),
respectively.}\label{Fig1}
\end{figure}

In Fig. 2, we show the partial density of states (PDOS)
corresponding to Ni 3$d$, Rh 4$d$ and O 2$p$ electronic states
obtained from GGA calculations, where Ni and Rh are
ferromagnetically coupled. The band splitting due to the crystal
field as discussed above is visible clearly in the figure. In Fig.
2(a), the energy range above -2 eV is essentially contributed by Ni
3$d$ electronic states. The lower energy region (-4 to -2 eV)
contains very low intensity of Ni 3$d$ PDOS and has dominant O 2$p$
contributions. The energy distributions of the density of states
(DOS) corresponding to both Ni 3$d$ and O 2$p$ states appear
similar. These results suggest that the DOS in -4 to -2 eV range can
be attributed to bonding states arising from Ni 3$d$ and O 2$p$
hybridizations and the antibonding states with dominant Ni 3$d$
character appear above -2 eV.

\begin{figure}
  \includegraphics[width=9cm]{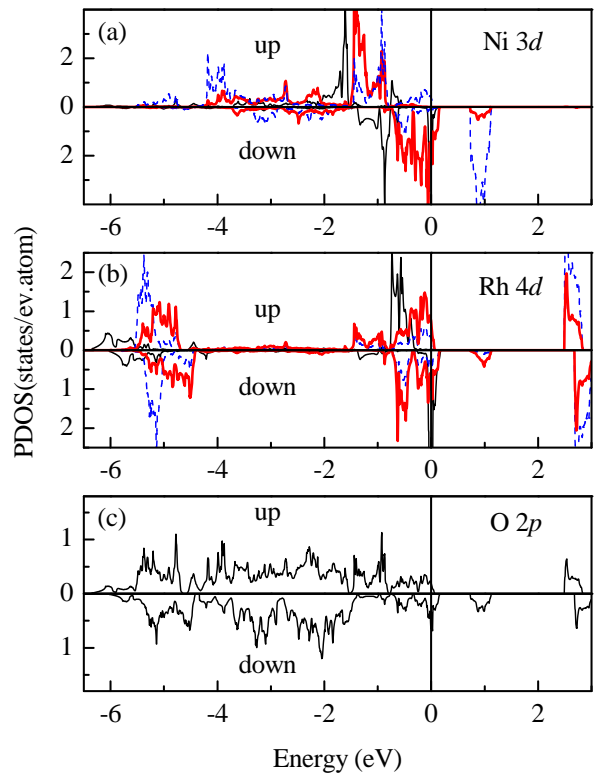}\\
  \caption{Color online) The partial density of states (PDOS)
corresponding to $d_0$ (thin solid lines), $d_{\pm1}$ (dashed lines)
and $d_{\pm2}$ (thick solid lines) orbitals for (a) Ni and (b) Rh
obtained from spin-polarized GGA calculation when Ni and Rh are
ferromagnetically coupled. (c) O 2$p$ PDOS.}\label{Fig2}
\end{figure}

The DOS in the energy ranges -6.3 to -4.4 eV and -1.5 to 0 eV
correspond to Rh 4$d$-O 2$p$ bonding and antibonding bands,
respectively. The intensity of Rh 4$d$ PDOS is almost similar in
these two regions, with negligible contribution in the energy region
of -4.4 to -1.5 eV (see Fig. 2(b)). Two distinct features are
observed in Fig. 2. (i) The energy separation between bonding and
antibonding bands arising due to Ni 3$d$-O 2$p$ hybridizations is
significantly smaller than the energy separation between the bonding
and antibonding Rh 4$d$-O 2$p$ bands. (ii) The bonding and
antibonding Ni 3$d$-O 2$p$ bands have dominant O 2$p$ and Ni 3$d$
character, respectively. However, the mixing of Rh 4$d$ and O 2$p$
characters in the Rh 4$d$-O 2$p$ hybridization is much stronger.
These observations clearly demonstrate the stronger O 2$p$-Rh 4$d$
hybridization presumably due to larger radial extension of the 4$d$
orbitals compared to the 3$d$ orbitals. Such large overlap integral,
$t$ leads to larger separation of the bonding and antibonding bands.
However, the bandwidth of the individual bands does not appear to
increase.

In addition, it is evident from Fig. 2(a) that up-spin channel of Ni
3$d$ states is almost fully occupied, whereas down-spin channel is
partially occupied. Major contribution in the unoccupied level comes
from $d_{\pm1}$ orbitals. In the case of Rh 4$d$ states, DOS
corresponding to both the spin states are partially occupied and
thus, contribute to the unoccupied DOS.

The GGA calculations for ferromagnetic (FM) intrachain coupling
converged to the metallic ground state in sharp contrast to the
experimentally observed insulating behavior in this compound.
Moreover, the total magnetic moment per formula unit (fu) obtained
from this calculation is about 2.85 $\mu_B$, which is much larger
than the experimentally estimated value of
$\sim$~1~$\mu_B$.\cite{niharika} The magnetic moment centered at Ni
and Rh sites are 1.55 $\mu_B$ and 0.43 $\mu_B$, respectively.
Magnetic moment centered at the oxygen sites is found to be 0.13
$\mu_B$, which is large and induced by the Ni 3$d$ and Rh 4$d$
moments.

It is evident from the magnetic moments that if Ni and Rh are
antiferromagnetically coupled, the total magnetic moment would match
with the experimental results. In order to investigate such
possibility, we have calculated the ground state energies and
wavefunctions corresponding to antiferromagnetic coupling between Ni
3$d$ and Rh 4$d$ moments. Interestingly, the calculated ground state
energy is found to be 27 meV/fu lower than that of FM state. The
calculated total magnetic moment comes out to be 1.04 $\mu_B$/fu.
Both these findings provide remarkable representation of the
experimental magnetization data.\cite{niharika} The calculated
magnetic moments for Ni and Rh are 1.31 $\mu_B$ and -0.36 $\mu_B$,
respectively. The moment centered at oxygen site is almost
negligible (0.02 $\mu_B$).

The calculated Ni 3$d$, Rh 4$d$ and O 2$p$ PDOS are shown in Fig. 3.
It is evident from the figure that intrachain antiferromagnetic
interaction leads to significant reduction in bandwidth of all the
energy bands and redistribution in spectral weight. For example, Ni
3$d$ up spin band is almost completely filled and the Fermi level,
$\epsilon_F$ is pinned at the top of the down spin band having
dominant $d_{\pm2}$ contributions and a width of about 0.5 eV while
Ni 3$d$ down spin band in Fig. 2 has a width of about 1 eV and is
partially filled. The occupancy of Rh 4$d$ up spin and down spin
PDOS exhibit spectral distribution opposite to that observed in Fig.
2 as expected due to antiferromagnetic coupling. In this case also
the bands are narrower than those observed in Fig. 2. The O 2$p$
contribution at the Fermi level is dominant for up spin while in the
ferromagnetic case it was down spin.

\begin{figure}
  \includegraphics[width=9cm]{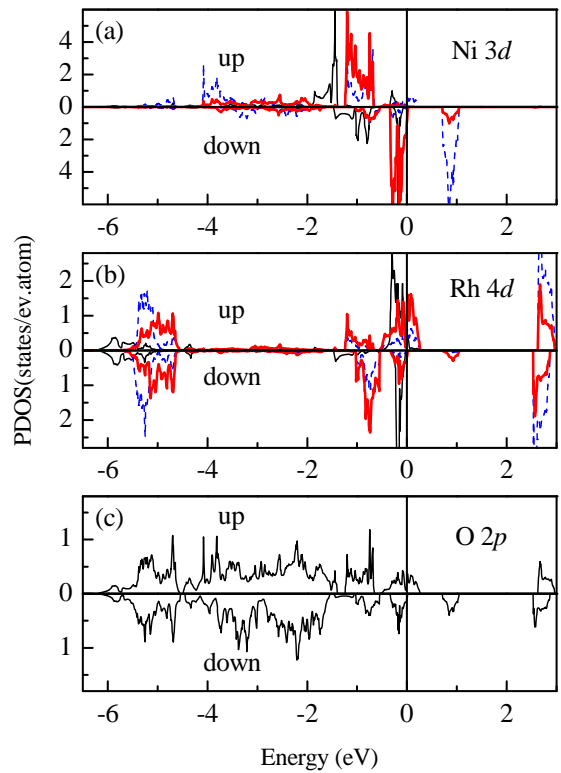}\\
  \caption{(Color online) The partial density of states (PDOS)
corresponding to $d_0$ (thin solid lines), $d_{\pm1}$ (dashed lines)
and $d_{\pm2}$ (thick solid lines) orbitals for (a) Ni and (b) Rh
obtained from spin-polarized GGA calculation when intrachain Ni and
Rh are antiferromagnetically coupled. (c) O 2$p$ PDOS.}\label{Fig3}
\end{figure}

All the above results clearly establish that the LSDA approach is
adequate to capture the magnetic phase in this system. Although,
there is significant spectral weight redistributions observed for
different magnetic configurations, both these calculations converge
to metallic ground state in contrast to the experimental studies.

It is well known that such {\it ab initio} calculations often
underestimate the on-site Coulomb correlation energy, which plays
significant role in determining the electronic properties of $d$ and
$f$ electron systems.\cite{anisimov97} The consideration of on-site
Coulomb interaction among the $d$ electrons ($U_{dd}$) under GGA+$U$
formulation is expected to improve the situation. Thus, we have
carried out GGA+$U$ calculations for different values of $U_{dd}$
corresponding to Ni 3$d$ electrons, which is defined as
$U_{dd}^{Ni}$ and Rh 4$d$ electrons, which is defined as
$U_{dd}^{Rh}$. The calculations were carried out considering the
intrachain antiferromagnetic coupling.

The calculated band gaps in the vicinity of $\epsilon_F$ for up spin
and down spin density of states are shown in Fig. 4. It is evident
from the figure that consideration of the electron correlation for
both Ni 3$d$ and Rh 4$d$ electrons helps to achieve the insulating
phase. We have shown two types of results in the figure, (i) the
dependence of the band gap as a function of $U_{dd}^{Ni}$ for a
fixed value of $U_{dd}^{Rh}$ = 3 eV and (ii) band gap as a function
of $U_{dd}^{Rh}$ for a fixed value of $U_{dd}^{Ni}$ = 5.5 eV. In the
up spin channel (see Fig. 4(a)), it is evident that an increase in
$U_{dd}^{Ni}$ does not lead to insulating phase till $U_{dd}^{Ni}$ =
4 eV. An increase in $U_{dd}^{Ni}$ above 4 eV provides an energy gap
of about 0.12 eV. Interestingly, the band gap remains unchanged for
further increase in $U_{dd}^{Ni}$. On the other hand, the
calculations as a function of $U_{dd}^{Rh}$ for $U_{dd}^{Ni}$ = 5.5
eV, indicate gradual increase in band gap. Thus, the energy gap is
essentially determined by $U_{dd}^{Rh}$ when $U_{dd}^{Ni}$ is kept
large. The Fig. 4(a) suggests that the creation of insulating ground
state requires $U_{dd}^{Rh}$ to be at least 2.6 eV. Since the
Coulomb repulsion strength depends inversely on the separation of
the two electrons, the strength of on-site Coulomb interaction is
sensitive to the radial extension of the wave functions. The present
result of $U_{dd}^{Ni}$ larger than $U_{dd}^{Rh}$ is consistent with
this behavior.

\begin{figure}
  \includegraphics[width=9cm]{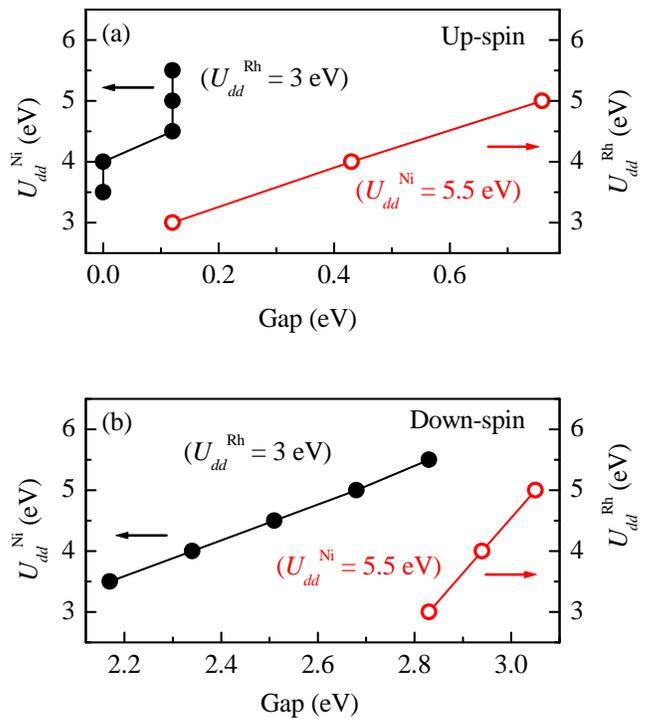}\\
  \caption{(Color online) On-site Coulomb interaction among Ni 3$d$
($U_{dd}^{Ni}$) and Rh 4$d$ ($U_{dd}^{Rh}$) electrons vs. insulating
gap for (a) up-spin and (b) down-spin channels. Two cases are
considered: (i) $U_{dd}^{Rh}$ is fixed at 3 eV and $U_{dd}^{Ni}$
varied from 3.5 to 5.5 eV (solid circles) and (ii) $U_{dd}^{Ni}$ is
fixed at 5.5 eV and $U_{dd}^{Rh}$ varied from 3 to 5 eV (open
circles).}\label{Fig4}
\end{figure}

Calculations for the down spin channel exhibit gradual increase in
band gap with the increase in $U_{dd}$ for both Ni 3$d$ and Rh 4$d$
electrons. The magnitude of the gap in this case is always much
higher ($>$ 2 eV) than that ($<$ 1 eV) observed for up spin density
of states. It is thus clear that various electronic properties of
this compound will essentially be controlled by electronic density
of states in the up spin channel.

In order to investigate the character of the electronic states in
the valence band, we show the calculated PDOS for $U_{dd}^{Ni}$ = 5
eV and $U_{dd}^{Rh}$ = 3 eV in Fig. 5. All the bands are
significantly modified due to the spectral weight transfer from
$\epsilon_F$ to the energy region away from it. Ni 3$d_{\pm1}$
contributions exhibit significant band narrowing. The up spin bands
appear around -5.8 eV and the down spin ones around 2.5 eV. The
contributions from other Ni 3$d$ orbitals appear between -2 eV to -5
eV. Interestingly, the consideration of $U_{dd}^{Ni}$ leads to a
significant increase of Ni 3$d$ character at higher energies. Thus
the O 2$p$ contributions relative to Ni 3$d$ become significantly
high in the vicinity of the Fermi level.

\begin{figure}
  \includegraphics[width=9cm]{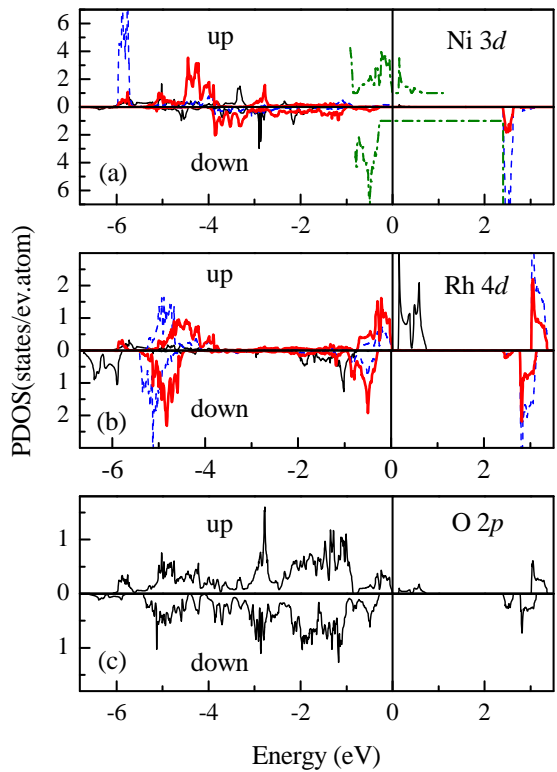}\\
  \caption{(Color online) The partial density of states (PDOS)
corresponding to $d_0$ (thin solid lines), $d_{\pm1}$ (dashed lines)
and $d_{\pm2}$ (thick solid lines) orbitals for (a) Ni and (b) Rh
obtained from spin-polarized GGA+$U$ calculation ($U_{dd}^{Ni}$ = 5
eV and $U_{dd}^{Rh}$ = 3 eV) when intrachain Ni and Rh are
antiferromagnetically coupled. (c) O 2$p$ PDOS.}\label{Fig5}
\end{figure}

Consideration of the on-site Coulomb interaction among Rh 4$d$
electrons reveals large effect on the density of states
distribution. In the GGA calculation $d_0$ was almost fully occupied
and there was a significant contribution of $d_{\pm1}$ and
$d_{\pm2}$ orbitals within 0.28 eV of $\epsilon_F$ in the unoccupied
up spin channel as evident from Fig. 3(b). This trend is reversed in
GGA+$U$ calculations, where region within 0.75 eV above the
$\epsilon_F$ is contributed by the electronic states having $d_0$
character. The electronic states below $\epsilon_F$ are contributed
by $d_{\pm1}$ and $d_{\pm2}$ orbitals.

Despite the fact that the Ni 3$d$ PDOS appear far away from
$\epsilon_F$, it has finite contribution in the vicinity of
$\epsilon_F$. This is evident in the rescaled Ni 3$d$ PDOS ($\times$
15 + 1) shown in Fig. 5(a) by dot-dashed line. Interestingly, the
energy distribution of these density of states closely follows the
distribution observed in Rh 4$d$ and O 2$p$ PDOS. This clearly
manifests the effect of hybridization between Ni 3$d$ and Rh 4$d$
states via O 2$p$ states. Most interestingly, the gap in the up spin
channel is much smaller than that found in down spin channel. Thus,
this system provides an unique example of a semiconductor, where the
electronic conduction is spin polarized. {\it While half-metallic
materials have long been investigated due to spin polarized
conduction of charge carriers in this system. This compound
manifests properties that can be more useful in the semiconductor
industry, where spin based technology is envisaged.}

Now, we discuss the manifestation of intrachain antiferromagnetic
coupling and Coulomb correlation in the band structure along
different high symmetry directions of the Brillouin zone. The
dispersion curves corresponding to different bands for FM, IAFM and
IAFM considering $U_{dd}^{Ni}$ = 5 and $U_{dd}^{Rh}$ = 3 eV for both
the spin channels are plotted in Fig. 6. The zero in the energy
scale indicates $\epsilon_F$. In the ferromagnetic solution, the up
spin bands along $\Gamma M$ and $\Gamma A$ directions do not cross
the Fermi level, while the down spin bands cross $\epsilon_F$ in all
the directions indicating half metallicity along these directions.
The antiferromagnetic coupling among Ni and Rh in the chains leads
to narrowing of the bands and the up spin bands cross the Fermi
level in all the directions. On the other hand the down spin bands
in the vicinity of $\epsilon_F$ essentially appear below
$\epsilon_F$. Again, the band crossing of the down spin band is
observed only along $K\Gamma$ direction suggesting half metallicity
in other directions.

\begin{figure}
  \includegraphics[width=9cm]{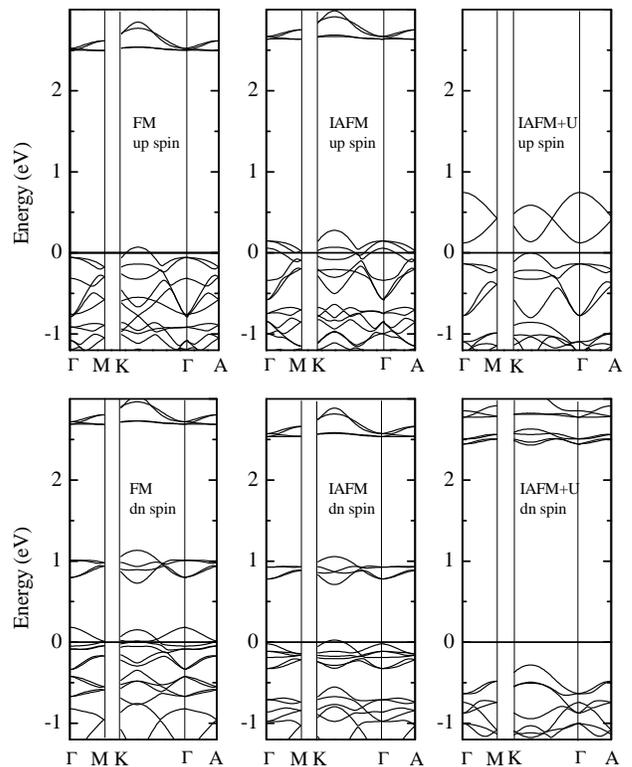}\\
  \caption{Dispersion curves along different symmetry directions of the
first Brillouin zone for both the spin channels. Fermi level is
denoted by zero in the energy scale.}\label{Fig6}
\end{figure}

Inclusion of the electron correlation among the $d$ electrons leads
to significant change in the band structure. While, there is a hard
gap in both the spin channels consistent with the transport behavior
of this compound, the gap exhibit significant $k$-dependence. The
gap at the high symmetry point, $\Gamma$ is lower than that observed
at $M$, $K$ or $A$ points. Interestingly, the dispersion of the
bands close to $\epsilon_F$ appears to be larger than that observed
in the uncorrelated case. This indicates that although electron
correlation essentially tends to localize the electrons, the
electronic states in the vicinity of the Fermi level are highly
mobile.

The occupancies of different $d$ orbitals corresponding to Ni and Rh
obtained from GGA+$U$ ($U_{dd}^{Ni}$ = 5 and $U_{dd}^{Rh}$ = 3 eV)
calculation are shown in Table 1. In up spin channel, all the Ni
3$d$ orbitals are almost completely filled and in down spin channel,
only $d_0$ orbital is completely filled. The small deviations
($\sim$ 0.05) from the completely filled values can be attributed to
the hybridization of Ni 3$d$ orbitals with the neighboring O 2$p$
orbitals. Total number of electron in the Ni 3$d$ is found to be
7.91, which suggests that the valency of Ni is close to 2+.

Table 1: The electron occupancies of $d_0$ , $d_{\pm1}$, and
$d_{\pm2}$ orbitals for Ni and Rh obtained from spin polarized
GGA+$U$ calculation. \vspace{2ex}

\begin{ruledtabular}
\begin{tabular}{|c|c|c|c|c|c|c|}

   & $d_{-2}$ & $d_{-1}$ & $d_0$ & $d_1$ & $d_2$ & Total \\ \hline
  Ni-up & 0.95 & 0.98 & 0.94 & 0.98 & 0.95 & 4.8 \\ \hline
  Ni-down & 0.8 & 0.28 & 0.95 & 0.28 & 0.8 & 3.11 \\ \hline
  Rh-up & 0.65 & 0.5 & 0.21 & 0.5 & 0.65 & 2.51 \\ \hline
  Rh-down & 0.66 & 0.52 & 0.82 & 0.52 & 0.66 & 3.18 \\ \hline

\end{tabular}
\end{ruledtabular}

In the case of Rh 4$d$ orbitals, both the spin channels are
partially occupied. Our calculations exhibit 5.69 electrons in the
Rh 4$d$ orbitals. If we consider O in 2- and Sr in 2+ states, then
charge neutrality condition demands Rh to be in 4+ state, which
corresponds to 5 electrons in the 4$d$ orbitals. The extra 0.69
electrons can be attributed to strong hybridization of Rh 4$d$
orbitals with the neighboring O 2$p$ orbitals.

The calculated magnetic moments for Ni and Rh ions are 1.69 $\mu_B$
and -0.68 $\mu_B$, respectively, with almost negligible contribution
from oxygen (0.01 $\mu_B$). The total magnetic moments per formula
unit, including the contribution of -0.07 $\mu_B$ from the
interstitial region, is found to be about 1 $\mu_B$ consistent with
the experimental results.\cite{niharika}

In order to understand the possible cause for IAFM interaction
between Ni and Rh spins, we show the schematic crystal field split
energy level diagram for both Ni and Rh obtained from spin
unpolarized GGA calculation by considering centers of gravity of
$d_0$, $d_{\pm1}$, and $d_{\pm2}$ PDOS is shown in Fig. 7. It is
evident from the figure that $d_0$ orbital has lowest energy for
both Ni and Rh. However, next energy levels consist of different
orbitals. In Ni, next level corresponds to $d_{\pm1}$ orbitals,
whereas in Rh it is $d_{\pm2}$ orbitals. The separation of Rh
$d_{\pm2}$ and $d_{\pm1}$ is found to be $\sim$0.7 eV, which is
about double of the Hund's coupling energy ($J_H$) of Rh 4$d$
electrons estimated from spin polarized GGA calculation. On the
other hand the energy separation of Ni $d_0$, $d_{\pm1}$, and
$d_{\pm2}$ orbitals are found to be $\sim$0.3 eV, which is about one
fifth of the estimated value of $J_H$ for Ni 3$d$ electrons.

\begin{figure}
  \includegraphics[width=9cm]{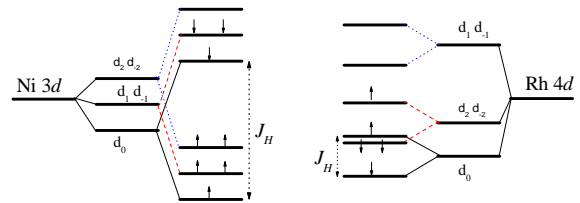}\\
  \vspace*{-5cm}
  \caption{(Color online) Schematic energy level diagram of $d_0$ ,
$d_{\pm1}$, and $d_{\pm2}$ orbitals for Ni$^{2+}$ and Rh$^{4+}$
showing antiferromagnetic coupling as the possible ground
state.}\label{Fig7}
\end{figure}

As discussed above, Ni is in 2+ state leading to $S$ = 1 state.
Rh$^{4+}$ has 5 electrons in the 4$d$ orbitals. The schematic
diagram clearly shows that Rh atoms will be in low-spin ($S$=1/2)
state with 2 electrons occupying $d_0$ orbital and the other 3
electrons will occupy $d_{\pm2}$ orbitals making them more than
half-filled. It is well known that in such a case, the interacting
electrons via oxygen 2$p$ orbitals (superexchange interaction) favor
antiparallel coupling in the ground state.\cite{goodenough}

\section{Conclusions}

In conclusion, we studied the electronic and magnetic properties of
Sr$_3$NiRhO$_6$ using {\it ab initio} spin polarized band structure
calculations within GGA and GGA+$U$. The calculations for intrachain
antiferromagnetic coupling exhibit the lowest energy. The magnetic
moments of Ni and Rh ions were found to be $\sim$1.7 $\mu_B$ and
-0.7 $\mu_B$, respectively, with a total moment of about 1 $\mu_B$.
All these results corresponding to the magnetic phase provide a
remarkable representation of the experimental results.

The insulating phase observed in transport measurements could not be
captured within GGA suggesting importance of electron correlation.
GGA+$U$ calculation reveals that the critical values of Coulomb
correlation strength to open up the insulating gap are $\sim$ 4.5
and $\sim$ 2.5 eV for Ni 3$d$ and Rh 4$d$ electrons, respectively.
The inclusion of electron correlation leads to significant change in
character of the energy bands. The feature close to $\epsilon_F$
exhibits enhancement of O 2$p$ character, while the $d$ character
enhances at higher energies. Interestingly, the band gap in up spin
channel is found to be significantly small ($\sim$ 0.12 eV) compared
to that observed ($>$ 2 eV) in the down spin channel indicating
possibilities of spin polarized conduction in the semiconducting
phase.

\section{Acknowledgements}

The authors acknowledge Prof. E.V. Sampathkumaran, TIFR, India for
drawing our attention towards this compound and useful discussions.


\begin{thebibliography}{99}

\bibitem{niitakajssc} S. Niitaka, H. Kageyama, M. Kato, K. Yoshimura, and
K. Kosuge, J. Solid State Chem. {\bf 146}, 137 (1999).

\bibitem{sampath} E. V. Sampathkumaran and A. Niazi, Phys. Rev. B {\bf 65},
180401(R) (2002).

\bibitem{flahaut} D. Flahaut, S. H$\acute{e}$bert, A. Maignan, V. Hardy,
C. Martin, M. Hervieu, M. Costes, B. Raquet, and J.M. Broto, Eur.
Phys. J B {\bf 35}, 317 (2003).

\bibitem{agrestini} S. Agrestini, C. Mazzoli, A. Bombardi, and M.R. Lees,
arXiv:0801.2495v1.

\bibitem{niitakaprl} S. Niitaka, K. Yoshimura, K. Kosuge, M. Nishi, and
K. Kakurai, Phys. Rev. Lett. {\bf 87}, 177202 (2001).

\bibitem{niharika} N. Mohapatra, K. K. Iyer, S. Rayaprol, and
E. V. Sampathkumaran, Phys. Rev. B {\bf 75}, 214422 (2007).

\bibitem{wuprb} H. Wu, Z. Hu, D. I. Khomskii, and L. H. Tjeng, Phys. Rev. B
{\bf 75}, 245118 (2007).

\bibitem{stitzer} K. E. Stitzer, W. H. Henley, J. B. Claridge, H. -C. zur Loye,
and R. C. Layland, J. Solid State Chem. {\bf 164}, 220 (2002).

\bibitem{sengupta} K. Sengupta, S. Rayaprol, K. K. Iyer, and E. V.
Sampathkumaran, Phys. Rev. B {\bf 68}, 012411 (2003).

\bibitem{whangbo} M. -H. Whangbo, D. Dai, H. -J. Koo, and S. Jobic,
Solid State Commun. {\bf 125} 413 (2003).

\bibitem{vidya} R. Vidya, P. Ravindran, H. Fjellv$\dot{a}$g, and A.
Kjekshus, Phys. Rev. Lett. {\bf 91}, 186404 (2003).

\bibitem{sampath1} E. V. Sampathkumaran, N. Fujiwara, S. Rayaprol, P. K. Madhu, and Y.
Uwatoko, Phys. Rev. B {\bf 70}, 014437 (2004).

\bibitem{fresard} R. Fr$\acute{e}$sard, C. Laschinger, T. Kopp, and V. Eyert,
Phys. Rev. B {\bf 69}, 140405(R) (2004).

\bibitem{wuprl} H. Wu, M. W. Haverkort, Z. Hu, D. I. Khomskii, and L. H. Tjeng,
Phys. Rev. Lett. {\bf 95}, 186401 (2005).

\bibitem{takubo} K. Takubo, T. Mizokawa, S. Hirata, J.-Y. Son,
A. Fujimori, D. Topwal, D. D. Sarma, S. Rayaprol, and E.-V. Sampathkumaran,
Phys. Rev. B {\bf 71}, 073406 (2005).

\bibitem{sugiyama} J. Sugiyama, H. Nozaki, Y. Ikedo, P. L. Russo, K. Mukai, D. Andreica,
A. Amato, T. Takami, and H. Ikuta, Phys. Rev. B {\bf 77}, 092409
(2008).

\bibitem{blaha} P. Blaha, K. Schwarz, G.K.H. Madsen, D. Kvasnicka, and J.
Luitz, WIEN2k, An Augmented Plane Wave + Local Orbitals Program for
Calculating Crystal Properties (Karlheinz Schwarz, Techn.
Universit$\ddot{a}$t Wien, Austria), 2001. ISBN 3-9501031-1-2.

\bibitem{wu} Z. Wu and R. E. Cohen, Phys. Rev. B {\bf 73}, 235116 (2006).

\bibitem{anisimov93} V. I. Anisimov, I. V. Solovyev, M. A. Korotin,
M. T. Czy$\dot{z}$yk and G. A. Sawatzky, Phys. Rev. B {\bf 48}, 16929 (1993).

\bibitem{anisimov97} V. I. Anisimov, F. Aryasetiawan, and A. I. Lichtenstein,
J. Phys.: Condens. Matter {\bf 9}, 767 (1997).

\bibitem{goodenough} J. B. Goodenough, in Magnetism and the Chemical Bond
(Interscience Publishers, New York-London, 1963).


\end{thebibliography}
\end{document}